\documentclass[a4paper,11pt]{article}
\pdfoutput=1 

\usepackage{jcappub} 

\usepackage[T1]{fontenc} 
\usepackage{deluxetable}
\usepackage{gensymb}
\usepackage{graphicx}
\usepackage{subcaption}
\usepackage{lipsum}
\usepackage{lscape}

\title{\boldmath Gamma-ray counterparts of radio astrophysical sources}

\author[a]{S. Best,}
\author[a,1]{J. Bazo \note{Corresponding author.}}

\affiliation[a]{Secci\'on F\'isica, Departamento de Ciencias, Pontificia Universidad Cat\'olica del Per\'u, Av. Universitaria 1801, Lima 32, Per\'u}

\emailAdd{sbest@pucp.edu.pe}
\emailAdd{jbazo@pucp.edu.pe}

\abstract{In this paper we study two newly discovered classes of radio sources: the highly energetic, short-lived events, known as Fast Radio Bursts (FRBs), and a new category of compact sources known as Fanaroff-Riley type 0 radio galaxies (FR0s). Due to a possible catastrophic event origin for the FRBs and a previous correlation found with an FR0 in the $\gamma$-ray spectrum, it is possible that these radio sources could also emit high energy photons in the Fermi-LAT satellite energy range (20 MeV - 300 GeV). Here we present an exhaustive time-dependent and spatial search of all up-to-date observed FRBs and FR0s, respectively. We perform a likelihood analysis of the radio sources by modeling the excess flux of gamma rays with a varying index power law function using data from Fermi-LAT and the 4FGL catalog. Sources with test statistic greater than 16 (corresponding to about 4$\sigma$) were further analyzed including 2 FRBs and 7 FR0s. No correlations with more than 5$\sigma$ were found after taking into account nearby sources. Therefore, upper limits for all sources were calculated.}

\begin{document}
\maketitle
\flushbottom

\section{Introduction} \label{sec:intro}

As more satellite and ground-based telescopes start to gather data covering different parts of the electromagnetic spectrum, new windows of opportunity for astronomy are opened. The possibility to observe the same source through different wavelengths can give us valuable information about the processes occurring inside our target. Although Fast Radio Bursts (FRBs) and Fanaroff-Riley type 0 galaxies (FR0) were first discovered in the radio spectrum, it is possible that these sources could also emit high energy photons in the Fermi-LAT satellite energy range (20 MeV - 300 GeV). This gamma-ray emission has been predicted, for example, in \citep{murase16} for FRBs. In addition, a FR0 (Tol1326-379) has been observed with Fermi \citep{grandi16}, whose SED has been modelled in \citep{tavecchio18}. 
Here we search for gamma-ray correlations from both types of radio source that could help us to better understand their origin and the engine driving their photon production.

Fast Radio Bursts (FRBs) are transient sources. They are very energetic, short lived (milliseconds), radio intense events which are thought to be extragalactic in origin due to their spatial distribution and their dispersion measure (DM) \citep{lorimer07}. Up to date, one repeating FRBs (FRB 121102 \citep{chatterjee17}) and two non-repeating FRBs (FRB 190523 \citep{ravi19} and FRB 180924 \citep{bannister19}) have been correlated with a known host galaxy. 

Some FRBs were found during searches for other astronomical events such as the Parkes Telescope search for galactic millisecond  pulsars, thus these FRBs are localized towards pulsars \citep{oslowski19}. However, these pulsars have different DM than the FRBs.

As the FRBs radio emissions already release a great amount of energy, if they were to emit gamma-rays as well, they could be candidates to most energetic events of the universe while providing important clues as to their origin. 

Possible origins for these gamma-ray emission signals are powerful events such as gamma-ray bursts \citep{zhang13} or neutron star merging \citep{totani13}. For a review of theoretical models explaining FRBs see \citep{katz18}. These gamma-ray emissions could occur before or after the radio emission itself and so we look for correlations before and after the arrival time of the radio signal.

The second source analyzed in this work are Fanaroff-Riley type 0 radio galaxies, which are steady sources. Fanaroff-Riley original classification \citep{fanaroff74} divides compact radio galaxies in type I (low intensity, with the bulk of the radio emissions coming from the core) and type II (high intensity with luminosity increasing in the lobes). Recently a new category, called type 0 \citep{baldi15}, has been added. FR0 galaxies are similar to type I except for a large deficit of extended radio structures. These are not believed to be FRIs seen from another angle due to FR0s being much more common \citep{baldi17}. 

A summary of 13 FRIs and 9 FRII detected with Fermi can be found at \citep{torresi18}, of which 6 FRIs have also a detected TeV counterpart. The study of these mis-aligned AGNs shows that there could be different emission regions of GeV photons, for instance, the expulsion of bright compact superluminal knots at sub-pc distance from the black hole or in extended radio lobes.

The study of FR0s in other wavelengths is crucial to better understand the engine driving the emission of these radio galaxies and to further differentiate them from FRIs.

In Section \ref{sec:obs} of this paper we briefly describe the historical radio observations leading to the discovery of these sources. Next, in Section \ref{sec:fermi} we detail the binned and unbinned likelihood analysis using Fermi-LAT data for both sources. In Section \ref{sec:results} we present the results of these analyses including the optimal set of parameters for the model of each source which maximizes the TS (Test Statistic) of the excess gamma-ray flux. Detailed information and further analysis of the sources with the highest TS ($\geq$ 16) are also presented. The significance of these correlations will be discussed as to offer a possible explanation.

\section{Radio Sources Observations} \label{sec:obs}

The sources mentioned in this paper (FRBs and FR0s) were both first detected in the radio spectrum. The first FRB was discovered in 2007 by Lorimer \citep{lorimer07}, but only recently new telescopes covering this range of frequencies have increased the number of detected FRBs. Now, with telescopes like the CHIME, we expect to see even dozens of FRBs each day, proving they are not isolated events. 

The FR0 class was added in 2015 to accommodate an increasingly number of sources which did not match well with the FRI class and which now constitute the most numerous class in the Fanaroff-Riley classification. 

\begin{figure}[ht!]
\centering
    \includegraphics[width=.6\textwidth]{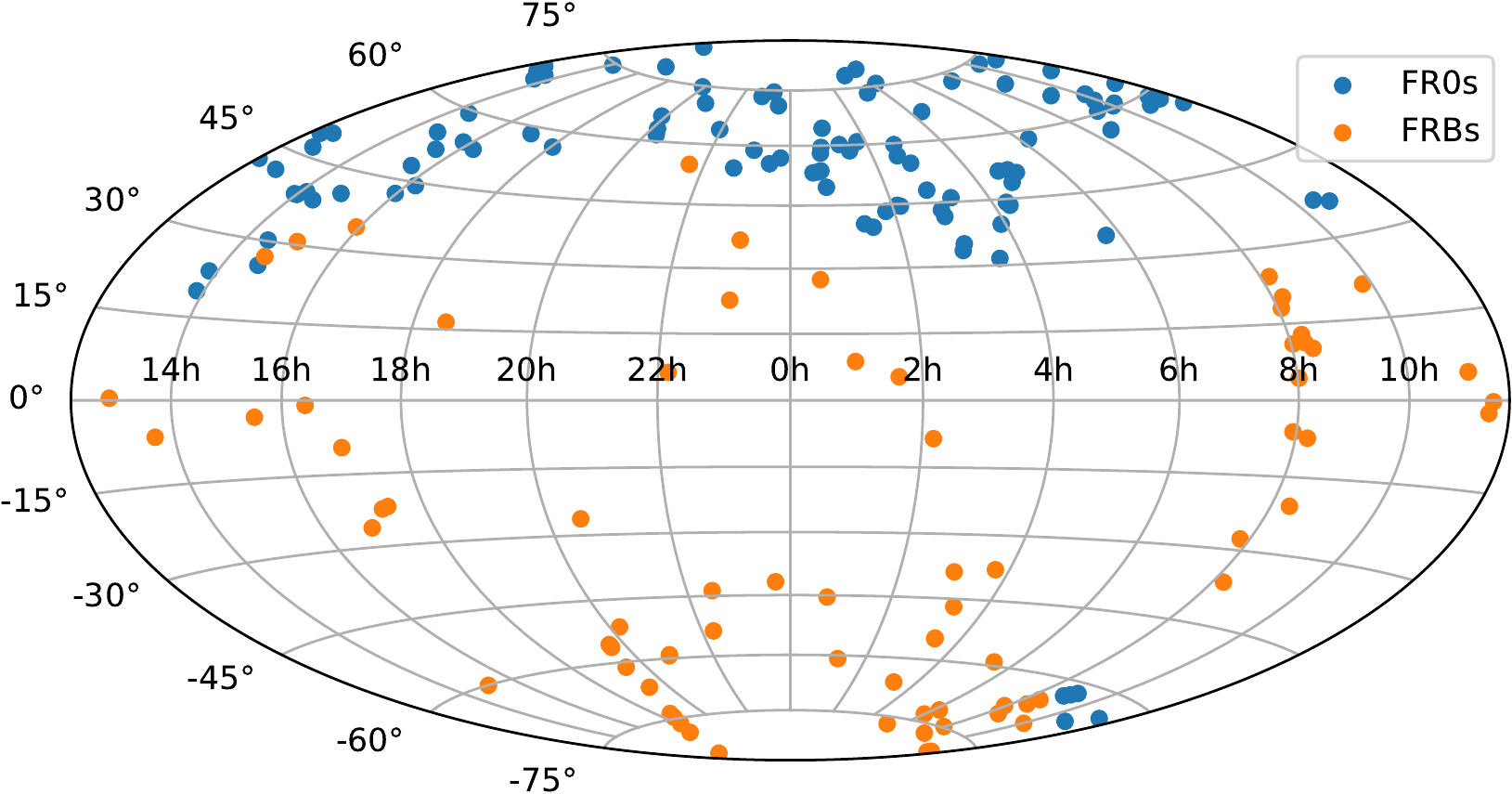}
    \caption{Sky map in galactic coordinates of all sources, both FRBs (from telescopes such as ASKAP, Parkes, UTMOST, all in Australia and CHIME in Canada) and FR0s (SDSS survey), studied in this paper. The spatial distribution of FRBs suggests their extragalactic origin.}
    \label{fig:map}
\end{figure}

We focus our attention for the FR0s (a very numerous source) in a subset of 108 sources catalogued \citep{baldi17} as part of the SSDS survey and in the case of FRBs, 76 events which had been detected by many telescopes (e.g. CHIME, ASKAP, Parkes, etc) \citep{petroff16} at the time of the analysis. A map in galactic coordinates of both types of sources can be seen in Fig. \ref{fig:map}, showing their extragalactic origin. 

There have been previous attempts to observe in other wavelengths both FRBs (X-rays \citep{scholz17} and $\gamma$-rays \cite{tendulkar16,yamasaki16,xi17}) and FR0s ($\gamma$-rays \citep{grandi16}). Nevertheless, the search (temporal in the case of FRBs and spatial in the case of FR0s) presented in this paper is more exhaustive, since it is applied to all sources listed in the available catalogs mentioned above. Furthermore, \cite{grandi16} starts with  Fermi's catalog association of 3FGLJ1330.0−3818 with the early-type galaxy Tol1326-379 and looks for alternative identifications of radio sources confirming that Tol1326-379 is the most likely
association. We instead look into Fermi data for candidates using the FR0s positions. As FR0s are constant in time, a temporal search is not necessary. For FRBs, a spatial search could also be performed, but this would increase the number of parameters being analyzed.

\section{Fermi-LAT data analysis} \label{sec:fermi}

The LAT, Large Area Telescope, on board of the Fermi Telescope satellite, is an imaging gamma-ray detector that uses pair production to detect both energy and momentum of high energy incident photons (20 MeV - 300 GeV). It covers, at any given moment, around 20\% of the sky with a resolution of arcminutes and has been collecting data since 2008 \citep{atwood09}.

We extracted the photon data files and spacecraft history from the LAT server using the time and coordinates provided by the FRB catalog \citep{petroff16} (76 sources) and the FR0 catalog \citep{baldi17} (108 sources) with an Astroquery \citep{astroquery} script. In the case of FRBs, we extracted data 25 hours around the radio event (1 day + 1 hour so the event is included in the central time bin). For FR0s we took the most recent year of data available at the moment of starting the analysis (from 3-9-18 to 3-9-19). Both included photons from 100 MeV to 300 GeV. For both studies we used Fermi LAT Pass 8 data with Fermi Science Tools cf201901. 

To reduce contamination from Earth's limb emission we used a maximum zenith angle (90$\degree$ for FR0s and 100$\degree$ for FRBs to increase the photon count) and took events only when the spacecraft was operating correctly ( [DAT $\&\&$ QUALITY] $ == $ 1 ). We then selected for events corresponding to all SOURCE class events (evclass$ = $128 and evtype$ = $3) and used the IRFs (Instrument Response Function): P8R3$\_$SOURCE$\_$V2. We took a ROI (region of interest) of 15$\degree$ and set the background sources using the 4FGL catalog \citep{fermi}. For convergence purposes, all source's index were fixed to their catalog value and only normalization was left free to vary up to 10$\degree$ from the center, after which normalizations were fixed. We used gll$\_$iem$\_$v06 to model the galactic background and iso$\_$P8R3$\_$SOURCE$\_$V2$\_$v1 for the extragalactic background.

For FRBs as well as for FR0s, we modeled the filtered data using a Power Law (Eq. \ref{eq:1}) with a varying $\Gamma$ index from 1.75 to 3.0 in steps of 0.25, with a fixed energy scale $E_0$ of 100 MeV. 

\begin{equation} \label{eq:1}
\dfrac{dN}{dE} = N_0\left(\dfrac{E}{E_0}\right)^{-\Gamma}
\end{equation}

We used a time-dependent search for the FRBs. Due to the low photon count, an    unbinned likelihood analysis was applied to each one hour bin, in which the photon data was divided. 
A spatial search, varying the source location by 0.1 $\degree$ in a grid of $1.1\degree\times1.1\degree$, was applied to the FR0s. In this case we used a binned likelihood analysis due to the large photon count for each source.

For each case we get a TS (i.e. likelihood-ratio test) value defined by: 

\begin{equation} \label{eq:3} 
TS = -2 log\left(\dfrac{L|\theta_0}{L|\theta_{ML}}\right) 
\end{equation}
where $L$ represents the likelihood given $\theta_0$, the set of parameters representing the null hypothesis, and $\theta_{ML}$, the most likely value of the set of parameter given the data. The square root of the TS can be treated as the detection significance, since we are fitting over only one parameter (i.e. $\theta_{0}=N_{0}$). 

Due to the large number of sources to be studied, we did the analysis described before as a preliminary one to select the most promising sources. Then, more detailed tests were used for the sources with higher TS ($\geq$ 16). For example, we applied an exponential cutoff model, given in Eq. \ref{eq:2}, extended the TS spatial map for FR0s, and analyzed their light curve and spectral energy distributions. In the case of the exponential cutoff model, we looped over both the additional $\Gamma_{2}$ indices, the energy cutoff, $E_C$, and left the normalization, $N_0$, to vary freely. $\Gamma$ was fixed to the value which gave the highest TS in the Power Law analysis. 

\begin{equation} \label{eq:2}
\dfrac{dN}{dE} = N_0\left(\dfrac{E}{E_0}\right)^{-\Gamma} \exp \left( - \left(\dfrac{E}{E_C}\right)^{-\Gamma_{2}}\right)
\end{equation}

\section{Results} \label{sec:results}

\begin{figure}[ht!]
\centering
  \includegraphics[width=.6\textwidth]{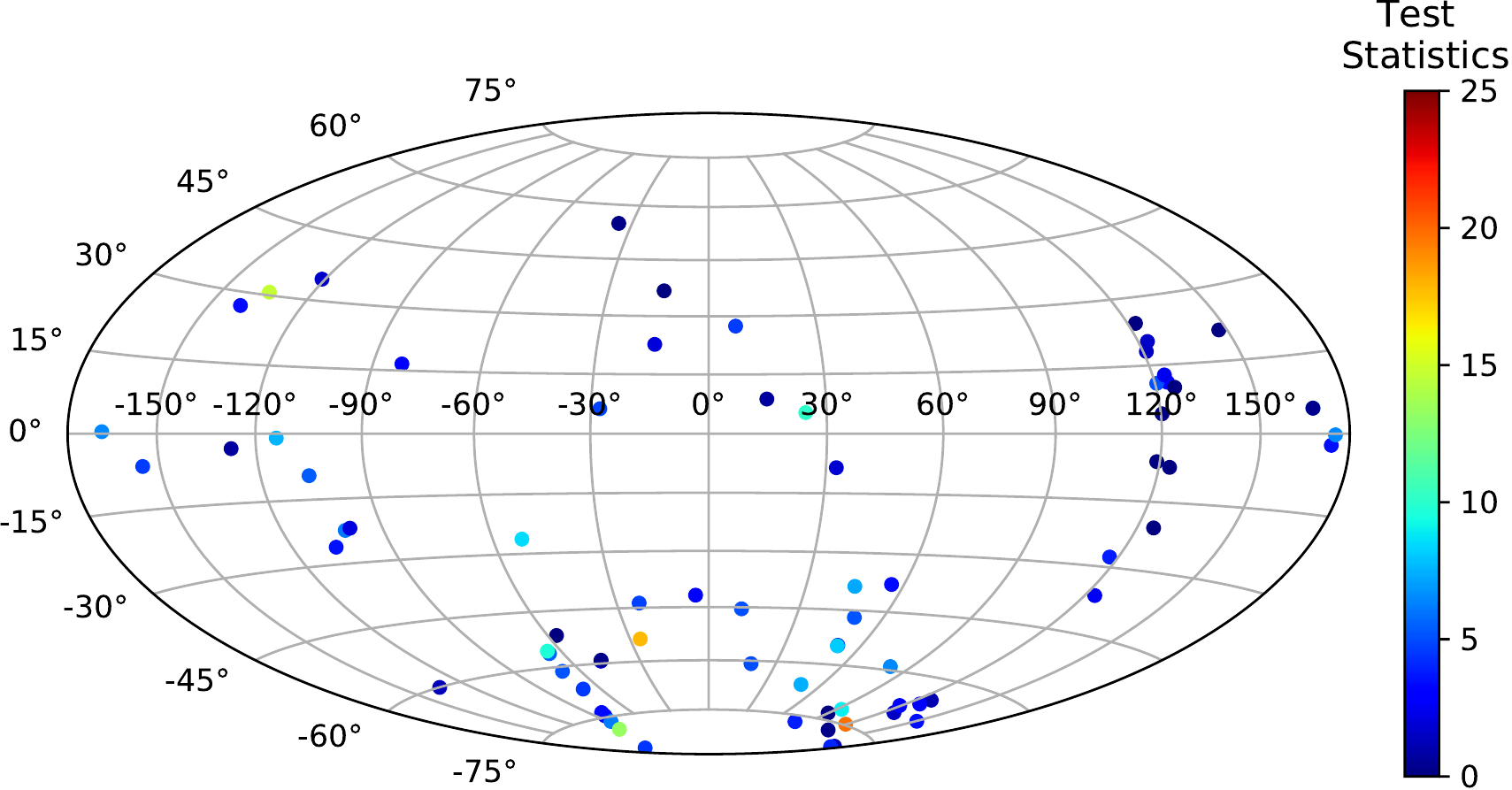}
    \caption{Sky map in galactic coordinates of 76 analyzed FRBs showing the maximum TS found according to the color scale. The TS corresponds to the best combination of time from the FRB event and $\Gamma$ index of the power law. The exact values are listed in Table \ref{tab:FRBs}.}
     \label{fig:FRB_TS_map}
\end{figure}

\begin{figure}[ht!]
\centering
 \includegraphics[width=.6\textwidth]{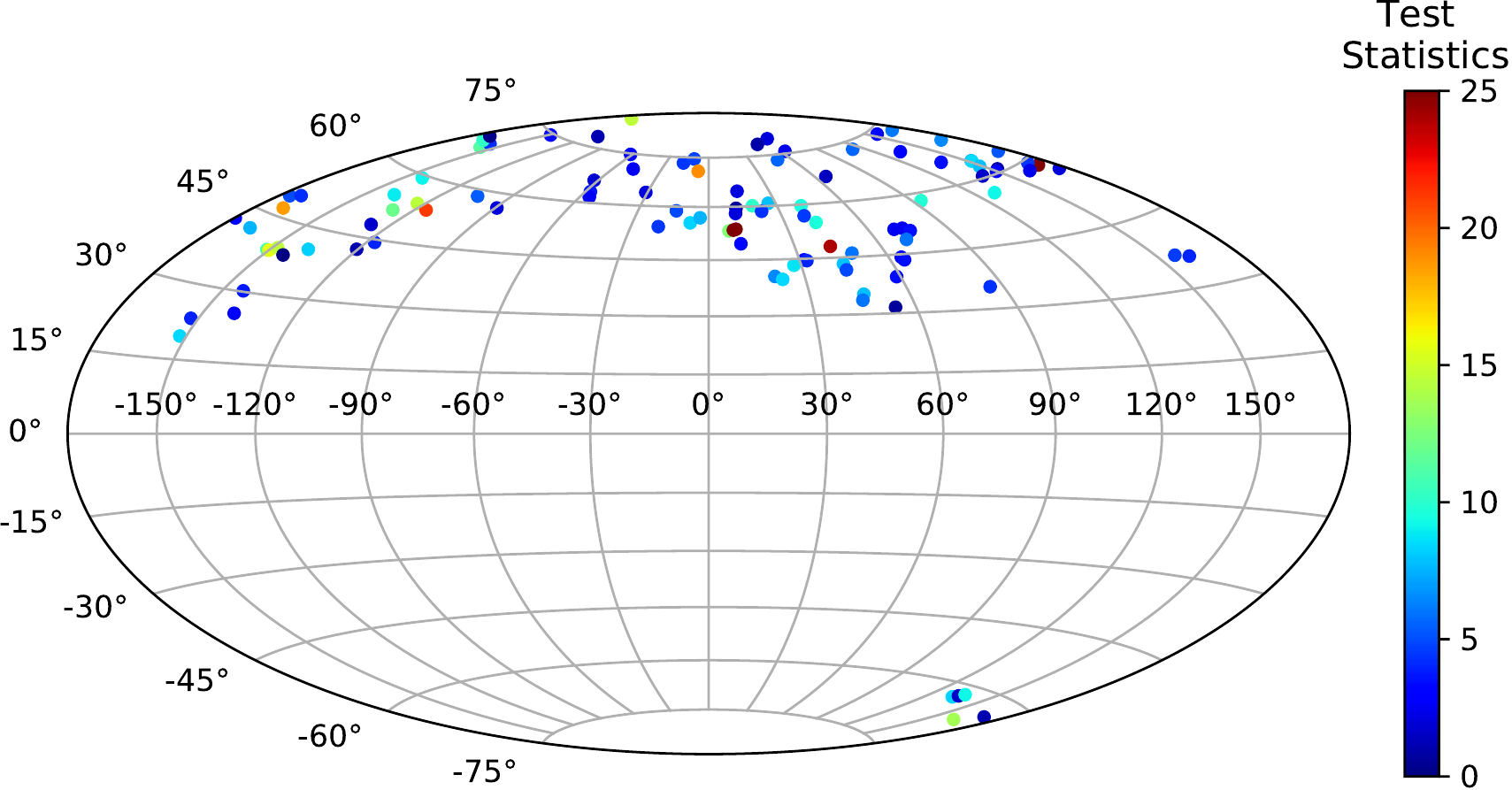}
    \caption{Sky map in galactic coordinates of 108 analyzed FR0s showing the maximum TS found according to the color scale. The TS corresponds to the best combination of the source's position in a spatial grid and $\Gamma$ index of the power law. The exact values are listed in Table \ref{tab:FR0s}.}
    \label{fig:FR0_TS_map}
\end{figure}

The results of our search can be found summarized in the Appendix in Table \ref{tab:FRBs} and Table \ref{tab:FR0s}, for FRBs and FR0s, respectively. We present for each source the set of parameters which maximizes the TS, including the $\Gamma$ index of the fitted power law (see Eq. \ref{eq:1}). In addition, in the case of FRBs, the time from the radio event is given, corresponding to the highest TS. For FR0s the distance from the detected radio source, in right ascension and declination, for the highest TS is shown. The resulting maximum TS is also presented as sky maps in galactic coordinates in Fig. \ref{fig:FRB_TS_map} and Fig.  \ref{fig:FR0_TS_map}, for FRBs and FR0s, respectively. 

We also give, for each source, the upper-limit of its luminosity ratio ($\gamma$ to radio) and the post-trial p-value including trials for the total number of sources of each type, the $\Gamma$ index steps, and the number of time and spatial bins (for FRBs and FR0s respectively). For the luminosity ratio calculation the $\gamma$-ray flux upper limit found is converted to energy flux integrating its spectral energy distribution function \ref{eq:1} using the found $\Gamma$ index. The radio spectral flux density (Jy) of FRBs is obtained from \cite{petroff16}. We assume a redshift of 0.5 (when no redshift was provided by the catalog) and a central frequency corresponding to the detecting telescope. The NVSS radio luminosity (erg s$^{-1}$) and redshift of FR0s is taken from \cite{baldi17}. In both cases we assume $H_0 = 67.8$ km s$^{-1}$ Mpc$^{-1}$.

In almost all cases we found a TS lower than 25 (approximately 5$\sigma$), so we calculated the flux upper limit with a 95\% confidence level, also shown in Tables \ref{tab:FRBs} and \ref{tab:FR0s}. There, we highlight the sources and parameters with a TS$\geq16$ which corresponds to roughly 4$\sigma$. There are only 2 FRBs (FRB150807 and FRB171004) and 7 FR0s (SDSSJ093003.56\-+341325.3, SDSSJ103719.33\-+433515.3,  SDSSJ104811.90\-+045954.8, SDSSJ135226.71\-+140528.5, SDSS\-J150601.89\-+084723.2, SDSSJ150636.57\-+092618.3 and SDSSJ155951.61\-+255626.3) satisfying this criteria. Given that these are the most promising sources, we give more details about them. 

\begin{figure}[ht!]
\centering
\begin{subfigure}{.49\textwidth}
 \centering
  \includegraphics[width=.99\linewidth]{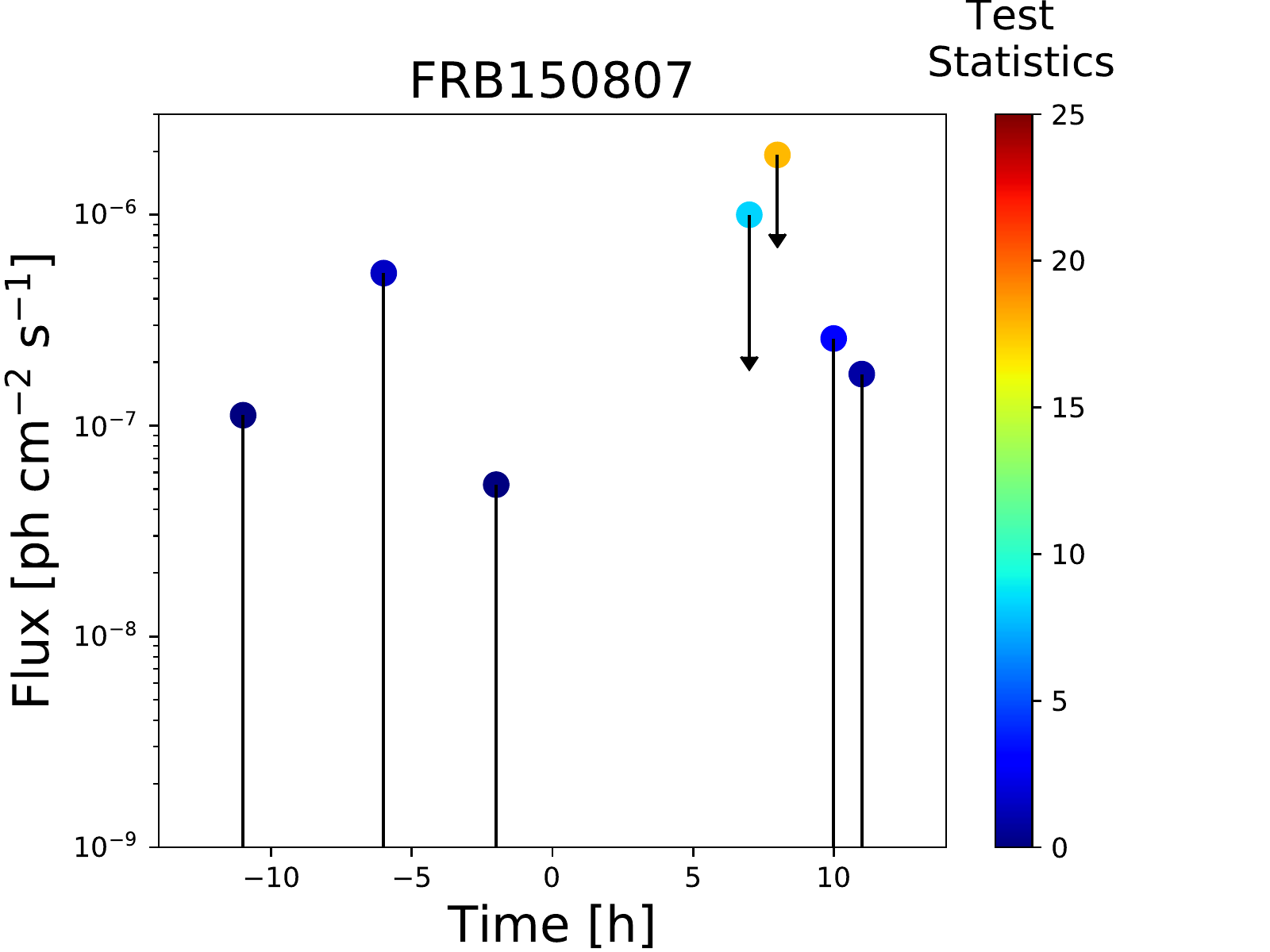}
  \end{subfigure}
    \begin{subfigure}{.49\textwidth}
  \centering
  \includegraphics[width=.99\linewidth]{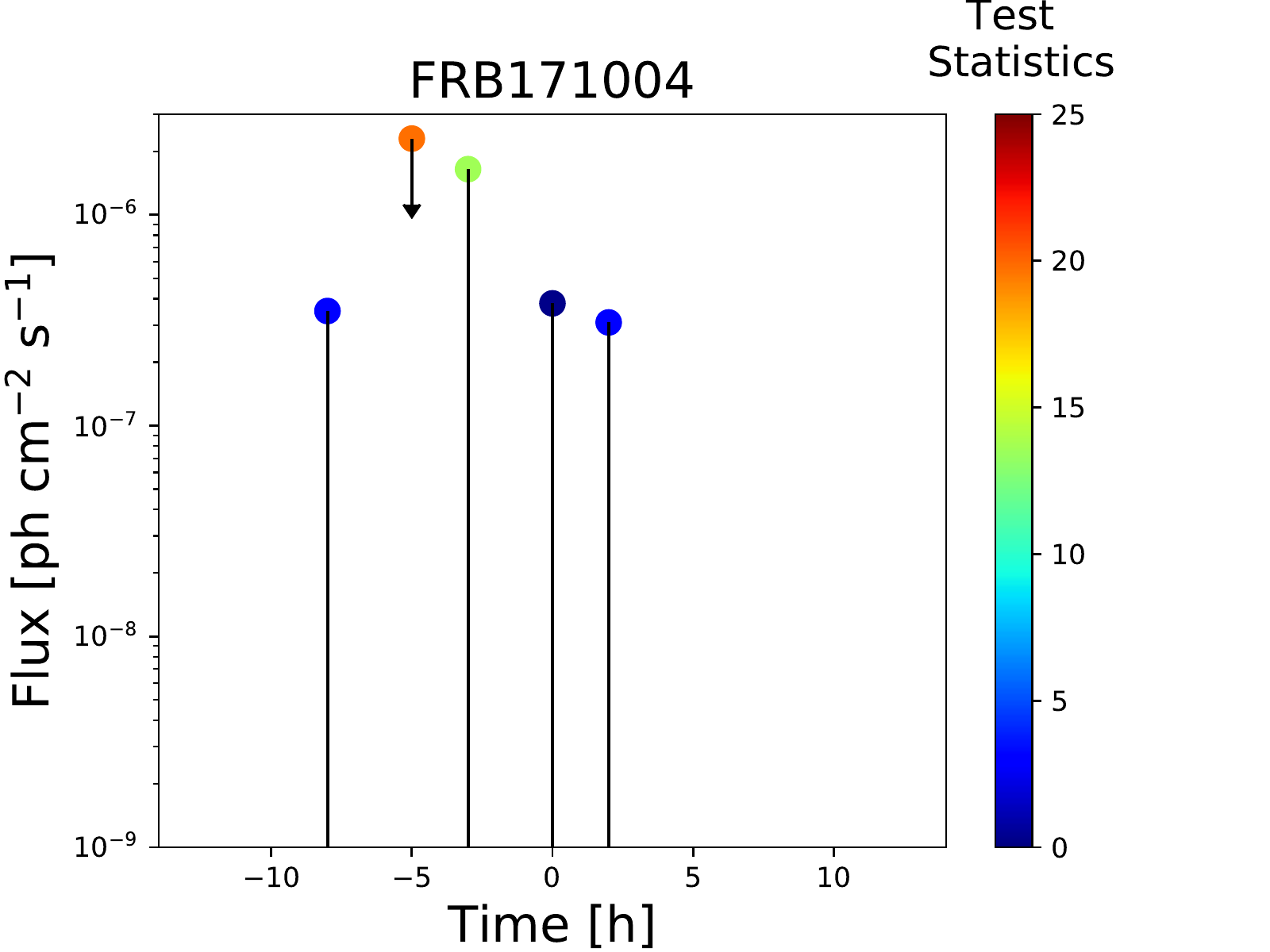}
   \end{subfigure}
    \caption{Light curves of the two most significant FRBs showing flux upper limits. The time axis is centered at the radio event's detection time using a 1 hour bin. The $\Gamma$ index corresponds to highest TS found and the color represents the TS. The missing points correspond to time bins without photon counts, thus the fit cannot converge. Left: FRB150807, max. TS = 17.84 ($\Gamma=3.0$, 8 hours after the radio event). Right: FRB171004, max. TS = 19.82 ($\Gamma=3.0$, 5 hours before the radio event).}
    \label{fig:FRBs}
\end{figure}

In Fig. \ref{fig:FRBs} we show the light curves of these two FRBs presenting flux upper limits. The upper limits are calculated using the $\Gamma$ index that gave the highest TS in one of the time bins. The markers' color represents the value of the TS. No significant  enhancement (i.e. flare) around the time of the event is observed. 

\begin{figure}[ht!]
\centering
  \includegraphics[width=.6\textwidth]
 {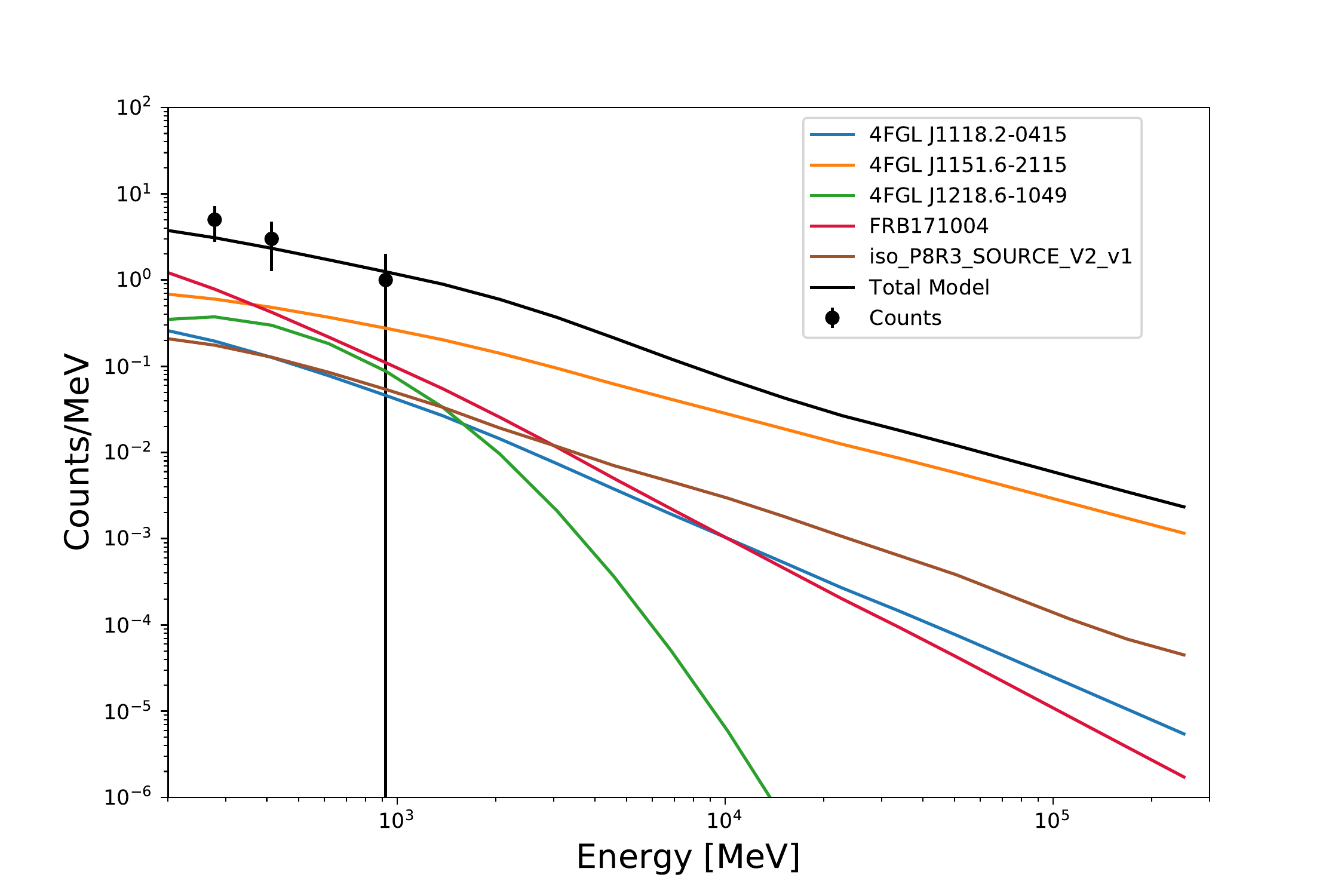}
    \caption{Spectral energy distribution of FRB171004 using the $\Gamma$ index ($\Gamma=3$) and time corresponding to the maximum TS. Points represent data from Fermi. Black line is the total model (i.e. the sum of the individual fits), red line represents the studied FRB. Other lines correspond to the most contributing sources.}
    \label{fig:SED_1}
\end{figure}

In addition, we analyzed the source's SED (spectral energy distribution) for both FRBs with TS$\geq16$. Since FRB150807 had just one data point (i.e. all photons fell into the same energy bin), we only present the SED of FRB171004 in Fig. \ref{fig:SED_1}. For low energies (few hundred MeV), this FRB contributes at the level of a nearby source to the total model. 

We then performed an analysis for both FRBs using an exponential cutoff (Eq. \ref{eq:2}) with $\Gamma_{2}$ index (from 1.0 to 2.5 in 0.5 steps), energy cutoff, $E_C$ (from $10^3$ to $10^6$ in 5 logarithmically spaced intervals), and left the normalization, $N_0$, to vary freely. $\Gamma$ was fixed to the value which gave the highest TS in the Power Law analysis. For both FRBs the results were less significant than with the Power Law model. For FRB150807 we get a TS of 15.89 and a flux upper limit of $1.93\times 10^{-6}$ ph cm$^{-2}$ s$^{-1}$ in the same time bin as before (8 hours after the radio event) with index $\Gamma_{2}=1.0$ and $E_C=10^{3.5}$ MeV. For FRB171004 we get a TS of 12.79 and a flux upper limit of $2.10 \times 10^{-6}$ ph cm$^{-2}$ s$^{-1}$ in the same time bin as before (5 hours before the radio event) with index $\Gamma_{2}=3.0$ and $E_C=10^3$ MeV.

\begin{figure}[ht!]
\begin{subfigure}{.49\textwidth}
 \centering
  \includegraphics[width=.99\linewidth]{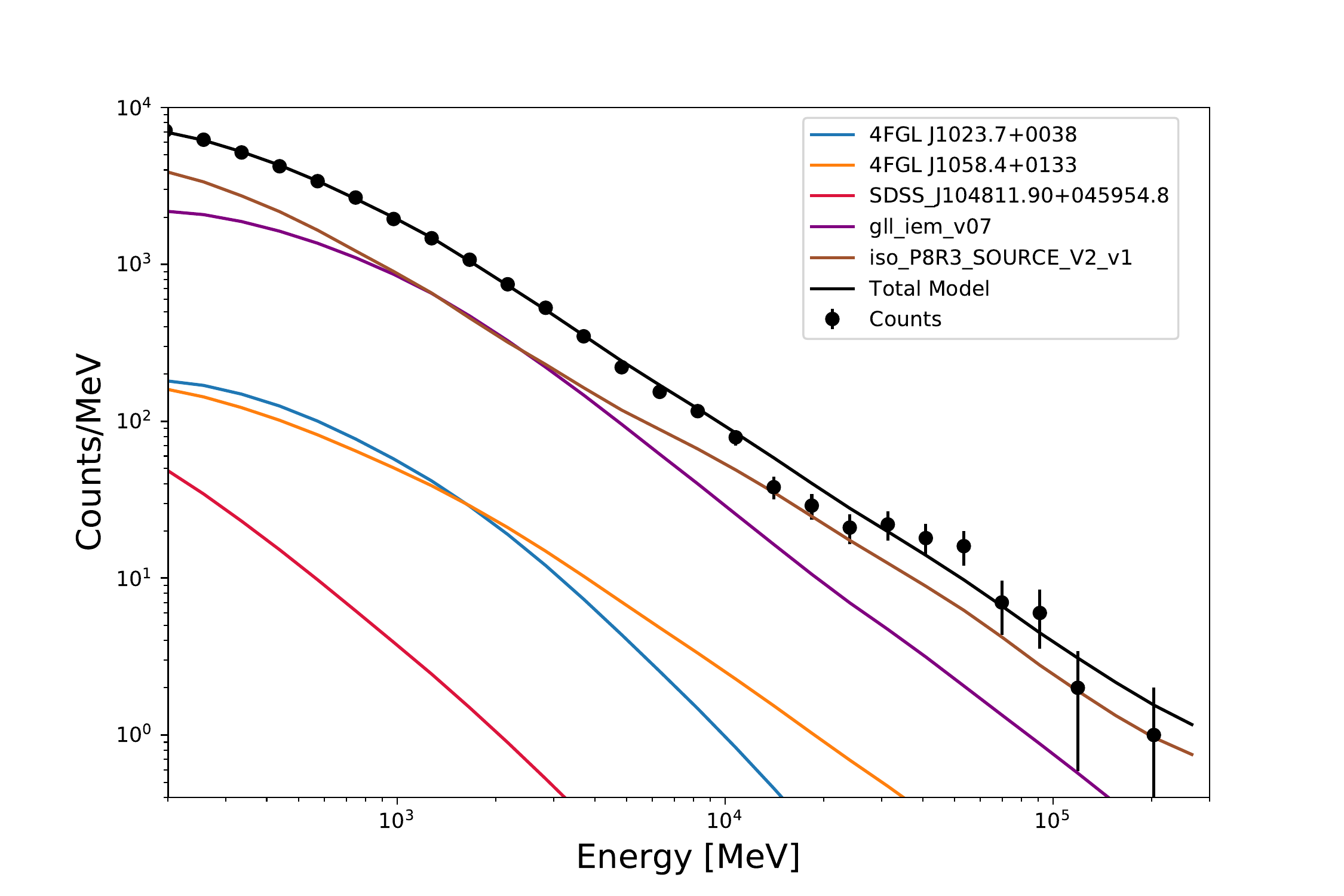}
  \end{subfigure}
    \begin{subfigure}{.49\textwidth}
  \centering
  \includegraphics[width=.99\linewidth]{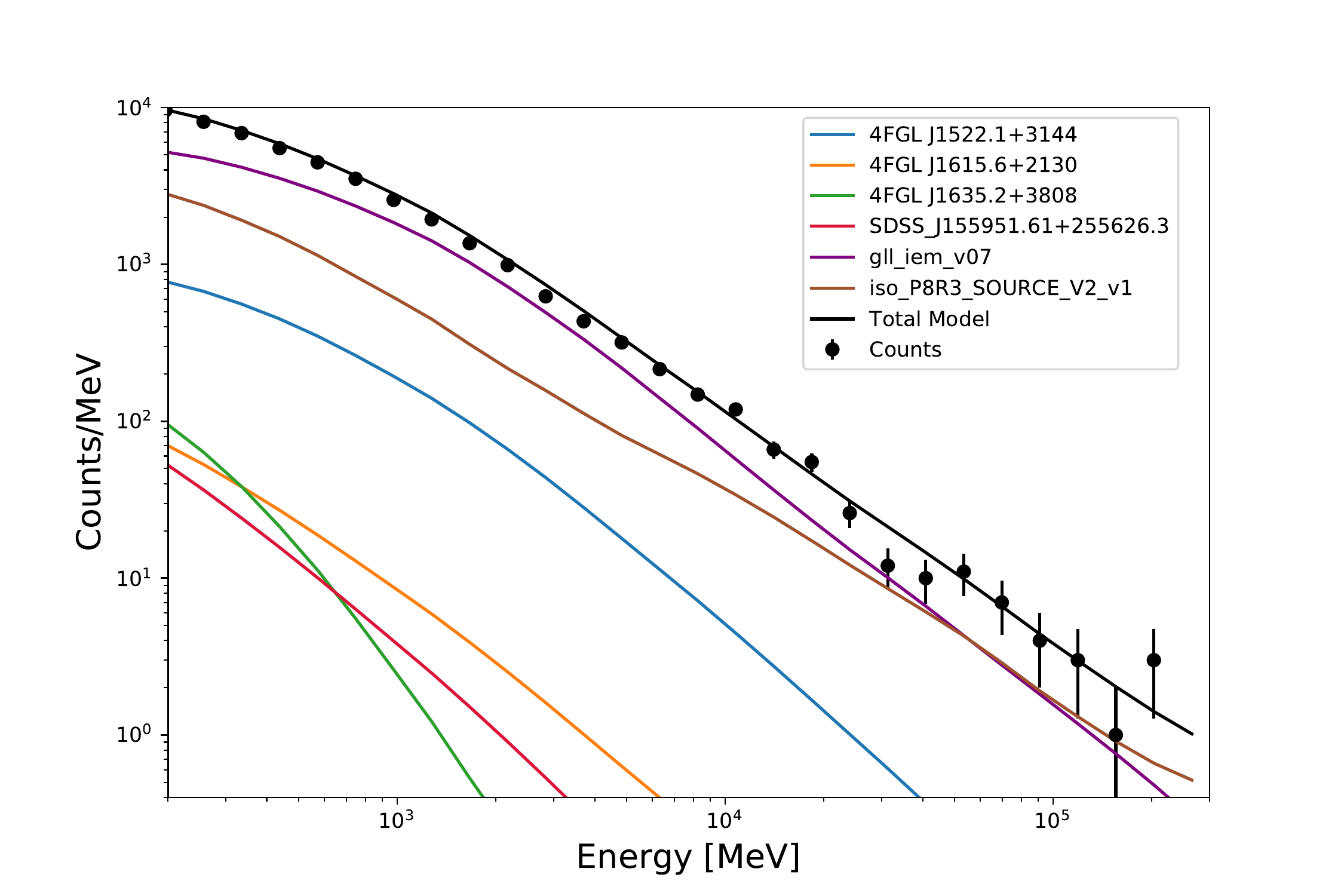}
   \end{subfigure}
    \caption{Spectral energy distribution of SDSS$\_$J104811.90+045954.8 (left) and SDSS$\_$J155951.61+255626.3 (right) using the $\Gamma$ index ($\Gamma=3$) and position corresponding to the maximum TS. Points represent data from Fermi. Black line is the total model (i.e. the sum of the individual fits), red line represents the studied sources. Other lines correspond to the most contributing sources.}
    \label{fig:FR0s_SED}
\end{figure}

Then we analyzed the seven FR0s with resulting TS$\geq16$. We show example SED plots (with the $\Gamma$ index and position of the apparent source corresponding to the highest TS) for two FR0s in Fig. \ref{fig:FR0s_SED}. As can be seen, the counts for these sources are much lower compared to nearby sources and the SED is actually dominated by the galactic and extragalactic backgrounds. This characteristic is shared by all 7 FR0s' SED.

We then contrasted the areas of high TS with known nearby sources from the 4FGL catalog. We found a possible origin to the high TS for 4 FR0s, shown in Table~\ref{tab:FR0_Close}. For the remaining 3 sources we checked the count map and could not find a possible origin. 

\begin{figure}[ht!]
\begin{subfigure}{.49\textwidth}
 \centering
  \includegraphics[width=.99\linewidth]{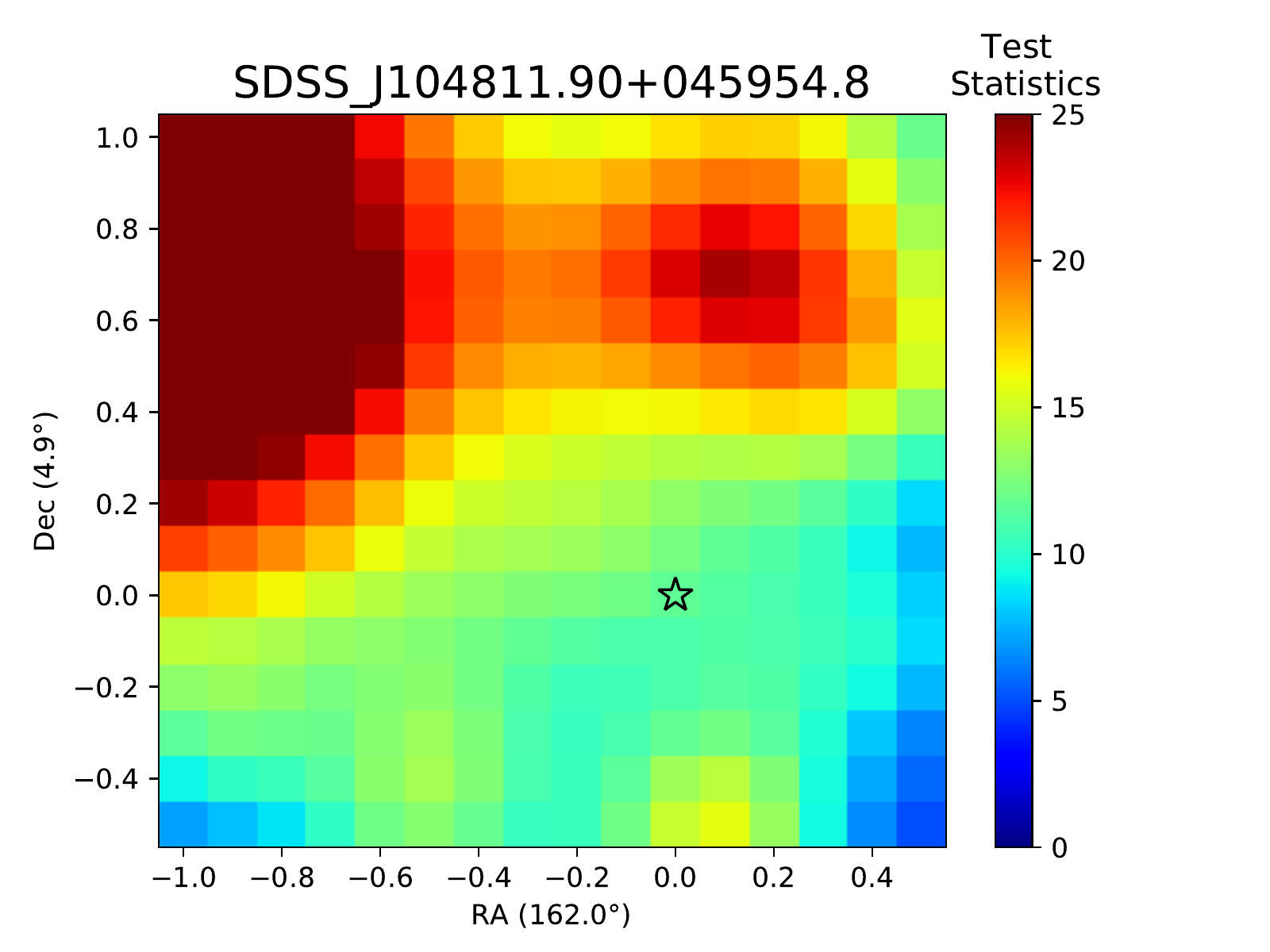}
  \end{subfigure}
    \begin{subfigure}{.49\textwidth}
  \centering
  \includegraphics[width=.99\linewidth]{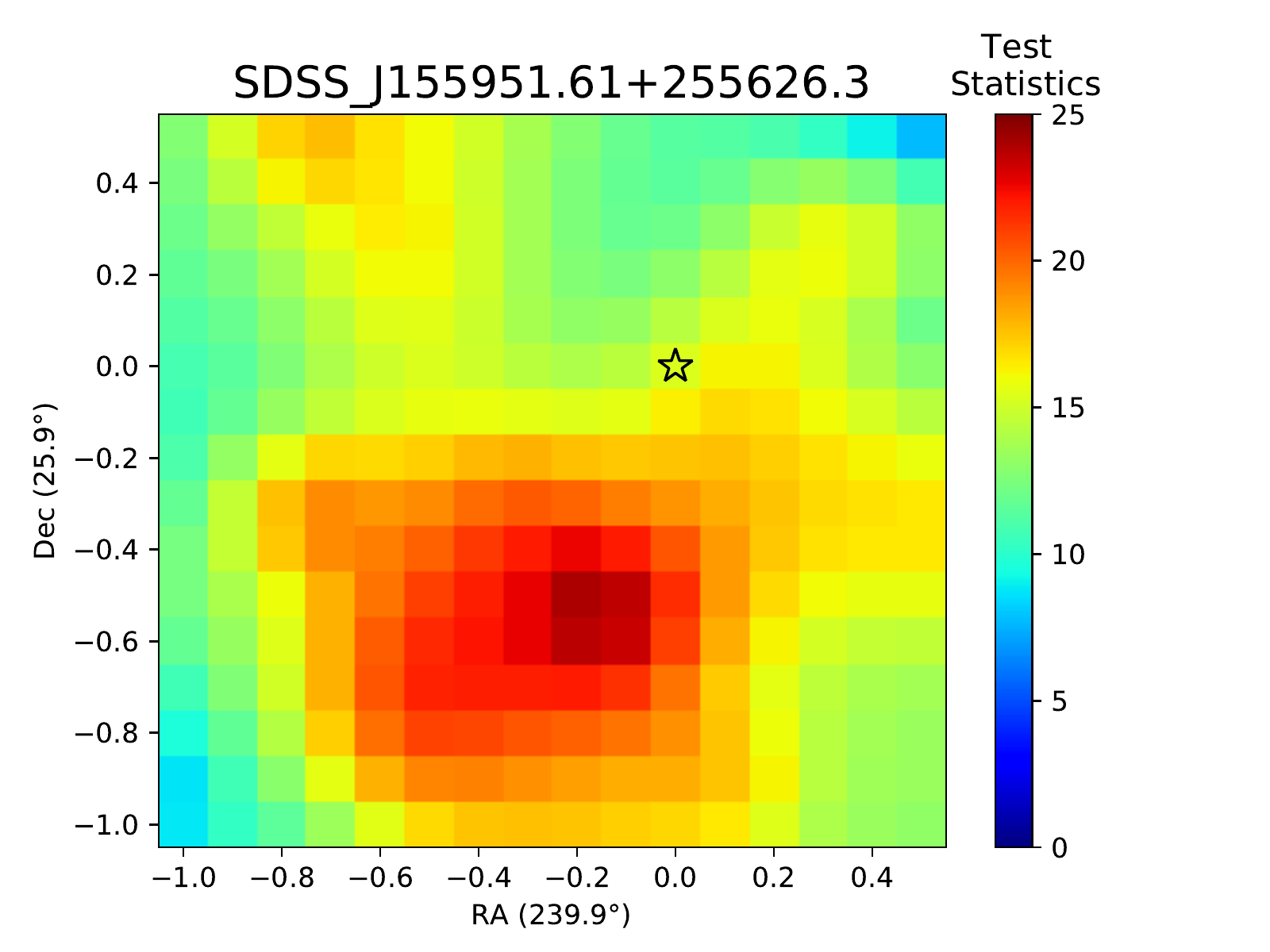}
   \end{subfigure}
    \caption{Test statistics extended spatial grid maps of Left: SDSS$\_$J104811.90+045954.8, Right: SDSS$\_$J155951.61+255626.3. Each pixel in the map represents a displacement of 0.1$\degree$ from the detected central position of the source, marked with a star, in right ascension and declination. The results correspond to the $\Gamma$ index that gave the maximum TS(Left: 3.0, Right: 3.0).}
    \label{fig:Expanded}
\end{figure}

The next step was to extend the TS Map (with the $\Gamma$ index corresponding to the highest TS) in the direction of the suspected high TS origin. The source position was varied from pixel to pixel by 0.1$\degree$ in a $1.5\degree\times1.5\degree$ grid, where the ($0\degree$,$0\degree$) position is the FR0 location, as described in the FR0 catalog.
In Fig. \ref{fig:Expanded} the apparent sources continue as expected for SDSS$\_$J104811.90\-+045954.8 and SDSS$\_$J155951.61\-+255626.3. 
For SDSS$\_$J135226.71+140528.5, however, the high TS region does not continue in the extended map (only two isolated points in the grid show high TS) which could be attributed to a statistical fluctuation and it is thus not shown here.\\

\begin{deluxetable}{lccccc}
\tablecaption{ Possible origin to the high TS for FR0s
\label{tab:FR0_Close}}
\tablecolumns{4}
\tablenum{1}
\tablewidth{0pt}
\tablehead{
\colhead{Source Name} &
\colhead{Nearby source}&
\colhead{R.A.} &
\colhead{Dec.} &
\colhead{Total} \\
\colhead{} & \colhead{} &
\colhead{dist. (\degree)}&\colhead{dist. (\degree)}&\colhead{dist. (\degree)}
}

\startdata
SDSS\_J093003.56+341325.3 & 4FGL J0930.7+3502&0.167&0.610&0.821\\
SDSS\_J103719.33+433515.3 & 4FGL J1035.6+4409&0.907&0.073&0.644\\
SDSS\_J150601.89+084723.2 & 4FGL J1504.4+1029&0.904&1.208&1.754\\
SDSS\_J150636.57+092618.3 & 4FGL J1504.4+1029&1.049&0.559&1.189\\
\enddata

\tablecomments{Column description: (1) Source name; Nearby source;(3) (4)  Right ascension (declination) (5) and Total distance.}
\end{deluxetable}

\section{Conclusions} \label{sec:conclusions}

We analyzed 76 FRBs and 108 FR0s, extragalactic sources discovered in radio, searching for a correlation with Fermi-LAT gamma ray data. We performed  time-dependent (FRB) and spatial (FR0) searches using an unbinned and binned likelihood analyses fixing the $\Gamma$ index of a power law that describes their flux, only letting one parameter (i.e. normalization) vary.

From these, 9 sources (two FRBs and seven FR0s) had a Test Statistic higher than 16 (roughly 4 $\sigma$). From the two FRBs we discarded FRB150807 due to only having one data point (i.e.: all photons fell in the same energy bin when making the fit). FRB171004 has a TS of 19.82 which is less than 5$\sigma$ to claim a correlation. After considering trial corrections the p-value reduces to 0.09, given the large number of FRBs, gamma index steps and time bins that were analyzed.

From the seven FR0s, 4 have nearby sources that could explain the high TS. From the remaining 3, SDSS$\_$J135226.71+140528.5 only appears to have high TS in two isolated points of the grid even when extended, so we attribute this to a statistical fluctuation. The remaining two FR0s (SDSS$\_$J104811.90+045954.8 and SDSS$\_$J155951.61+255626.3) could be discarded due to their SED, since in both cases the background is dominant. This absence of a $\gamma$-ray signal by FR0s is in agreement with expectations that these objects will be too faint in gamma-rays to be detected individually by Fermi \citep{stecker19}.

For FRBs we found Fermi gamma luminosity upper limits (95\% C.L.) in the range $10^{46} \lesssim L_{\gamma} (\mbox{erg~s}^{-1}) \lesssim 10^{48}$. This gives us luminosity ratios ($L_{\gamma}/L_{radio}$) in the order of $10^{4}$. The same ratio order is found for FR0s for which $10^{41} \lesssim L_{\gamma} (\mbox{erg~s}^{-1}) \lesssim 10^{43}$. In the case of FR0 Tol1326-379, which was found in $\gamma$-rays, the measured luminosity ratio found in \citep{grandi16} is L$_\gamma$/L$_{radio}$ $\sim 1\times10^3$, using 4 years of Fermi data (3LAC). We can set the following upper limits for the most promising FR0 sources in the catalog: SDSS$\_$J104811.90+045954.8 (L$_\gamma$/L$_{radio}$ $\sim 1\times10^4$) and SDSS$\_$J155951.61+255626.3 (L$_\gamma$/L$_{radio}$ $\sim 2.9\times10^3$).

\appendix
\section{Extended Results Tables}

\begin{landscape}
\begin{deluxetable}{ccccccc}
\tablecaption{Fast Radio Bursts: summary of the parameters for the maximum TS.
\label{tab:FRBs}}
\tablecolumns{7}
\tablenum{1}
\tablewidth{0pt}
\tablehead{
\colhead{Source} &
\colhead{Gamma} & 
\colhead{Time from} &
\colhead{Flux (10$^{-6}$)} & 
\colhead{Test}&
\colhead{L$_\gamma$/L$_{radio}$}&
\colhead{p-value} \\
\colhead{Name} & 
\colhead{Index ($\Gamma$)} &
\colhead{event (h)} &
\colhead{(ph cm$^{-2}$ s$^{-1}$)} &
\colhead{Statistic}&
\colhead{(10$^{4}$)}&
\colhead{(post trial)}
}

\startdata
FRB090625 & 2.0 & 0.0 & 0.15 & 1.32 & 1.27 & 1.00 \\
FRB110214 & 1.75 & -10.0 & 0.21 & 4.57 & 0.18 & 1.00 \\
FRB110220 & 1.75 & -12.0 & 0.25 & 8.19 & 4.35 & 1.00 \\
FRB110523 & 3.0 & 10.0 & 1.02 & 3.39 & 3.97 & 1.00 \\
FRB110626 & 2.75 & -8.0 & 0.46 & 2.82 & 3.10 & 1.00 \\
FRB110703 & 2.75 & 7.0 & 0.64 & 6.54 & 3.46 & 1.00 \\
FRB120127 & 2.25 & 7.0 & 0.98 & 7.51 & 9.91 & 1.00 \\
FRB121002 & 2.25 & -5.0 & 0.83 & 8.59 & 9.72 & 1.00 \\
FRB121102 & 3.0 & -7.0 & 3.67 & 6.51 & 21.39 & 1.00 \\
FRB130626 & 3.0 & -9.0 & 1.03 & 4.66 & 3.23 & 1.00 \\
FRB130628 & 1.75 & -3.0 & 0.92 & 14.68 & 10.79 & 0.77 \\
FRB130729 & 3.0 & 4.0 & 0.20 & 0.23 & 2.09 & 1.00 \\
FRB131104 & 3.0 & 2.0 & 0.27 & 2.10 & 0.57 & 1.00 \\
FRB140514 & 3.0 & 9.0 & 0.26 & 2.04 & 1.31 & 1.00 \\
FRB141113 & 2.25 & 0.0 & 3.26 & 6.51 & 421.68 & 1.00 \\
FRB150215 & 2.75 & 0.0 & 4.16 & 10.10 & 16.13 & 1.00 \\
FRB150418 & 2.0 & 9.0 & 0.54 & 0.74 & 2.30 & 1.00 \\
FRB150610 & 2.5 & -12.0 & 0.23 & 2.84 & 1.13 & 1.00 \\
\textbf{ FRB150807 } & \textbf{ 3.0 } & \textbf{ 8.0 } & \textbf{ 1.92 } & \textbf{ 17.84 } & \textbf{ 0.04 } & \textbf{ 0.24 }\\
FRB151018 & 1.75 & -9.0 & 0.20 & 3.50 & 3.24 & 1.00 \\
FRB151206 & 2.75 & 1.0 & 1.24 & 1.85 & 11.20 & 1.00 \\
FRB151230 & 2.75 & 12.0 & 0.58 & 1.60 & 3.76 & 1.00 \\
FRB160102 & 2.25 & 9.0 & 1.04 & 5.08 & 10.48 & 1.00 \\
FRB160317 & 2.5 & 5.0 & 3.88 & 7.61 & 4.44 & 1.00 \\
FRB160410 & 3.0 & 12.0 & 0.56 & 3.36 & 0.19 & 1.00 \\
FRB160608 & 3.0 & -9.0 & 0.84 & 5.46 & 0.45 & 1.00 \\
FRB160920 & 3.0 & -7.0 & 0.30 & 0.33 & 3.14 & 1.00 \\
FRB170107 & 1.75 & 6.0 & 0.21 & 3.16 & 0.19 & 1.00 \\
FRB170416 & 2.5 & -11.0 & 1.49 & 5.25 & 0.26 & 1.00 \\
FRB170428 & 2.25 & -9.0 & 0.76 & 3.38 & 0.50 & 1.00 \\
FRB170606 & 3.0 & -8.0 & 8.41 & 6.26 & 36.36 & 1.00 \\
FRB170707 & 3.0 & 11.0 & 0.10 & 0.28 & 0.02 & 1.00 \\
FRB170712 & 3.0 & 8.0 & 1.06 & 9.71 & 0.07 & 1.00 \\
FRB170827 & 1.75 & -8.0 & 0.00 & 0.00 & 0.00 & 1.00 \\
FRB170906 & 3.0 & 4.0 & 1.12 & 4.09 & 0.09 & 1.00 \\
FRB170922 & 2.5 & -1.0 & 2.03 & 7.32 & 1.35 & 1.00 \\
FRB171003 & 2.75 & -7.0 & 0.36 & 0.20 & 0.02 & 1.00 \\
\textbf{ FRB171004 } & \textbf{ 3.0 } & \textbf{ -5.0 } & \textbf{ 2.30 } & \textbf{ 19.82 } & \textbf{ 0.24 } & \textbf{ 0.09 }\\
FRB171019 & 2.0 & 0.0 & 0.46 & 3.48 & 0.11 & 1.00 \\
FRB171020 & 3.0 & 5.0 & 0.29 & 1.30 & 0.01 & 1.00 \\
FRB171116 & 1.75 & 10.0 & 0.35 & 3.84 & 0.40 & 1.00 \\
FRB171209 & 1.75 & -1.0 & 1.00 & 4.80 & 15.23 & 1.00 \\
FRB171213 & 2.0 & 4.0 & 0.32 & 9.06 & 0.03 & 1.00 \\
FRB171216 & 3.0 & 1.0 & 0.22 & 2.45 & 0.02 & 1.00 \\
FRB180110 & 2.75 & -5.0 & 0.87 & 13.35 & 0.02 & 0.95 \\
FRB180119 & 1.75 & -12.0 & 0.00 & 0.00 & 0.00 & 1.00 \\
FRB180128.0 & 3.0 & -3.0 & 0.89 & 1.51 & 0.12 & 1.00 \\
FRB180128.2 & 2.5 & 7.0 & 0.47 & 5.53 & 0.06 & 1.00 \\
FRB180130 & 3.0 & -10.0 & 0.79 & 6.16 & 0.08 & 1.00 \\
FRB180131 & 2.75 & 10.0 & 0.34 & 3.05 & 0.04 & 1.00 \\
FRB180212 & 3.0 & -12.0 & 0.88 & 3.19 & 0.04 & 1.00 \\
FRB180301 & 3.0 & -4.0 & 10.88 & 4.62 & * & 1.00 \\
FRB180309 & 3.0 & -10.0 & 0.92 & 5.10 & 0.08 & 1.00 \\
FRB180311 & 3.0 & 2.0 & 0.57 & 4.77 & 8.91 & 1.00 \\
FRB180315 & 2.0 & -10.0 & 0.60 & 4.42 & 0.24 & 1.00 \\
FRB180525 & 2.0 & 3.0 & 0.65 & 1.14 & 0.08 & 1.00 \\
FRB180528 & 3.0 & -11.0 & 1.80 & 6.10 & 0.27 & 1.00 \\
FRB180714 & 3.0 & 5.0 & 3.27 & 0.71 & 12.73 & 1.00 \\
FRB180725.J0613+67 & 1.75 & -11.0 & 0.00 & 0.00 & * & 1.00 \\
FRB180725A & 1.75 & -11.0 & 0.00 & 0.00 & * & 1.00 \\
FRB180727.J1311+26 & 2.25 & -12.0 & 0.48 & 2.60 & * & 1.00 \\
FRB180729.J0558+56 & 1.75 & -12.0 & 0.00 & 0.00 & * & 1.00 \\
FRB180729.J1316+55 & 1.75 & -12.0 & 0.00 & 0.00 & * & 1.00 \\
FRB180730.J0353+87 & 1.75 & -12.0 & 0.00 & 0.00 & * & 1.00 \\
FRB180801.J2130+72 & 1.75 & -12.0 & 0.00 & 0.00 & * & 1.00 \\
FRB180806.J1515+75 & 2.25 & -10.0 & 0.19 & 2.51 & * & 1.00 \\
FRB180810.J0646+34 & 2.75 & 12.0 & 0.87 & 3.83 & * & 1.00 \\
FRB180810.J1159+83 & 3.0 & 9.0 & 0.31 & 1.61 & * & 1.00 \\
FRB180812.J0112+80 & 2.0 & 2.0 & 0.21 & 1.51 & * & 1.00 \\
FRB180814.J0422+73 & 3.0 & -10.0 & 0.60 & 5.14 & * & 1.00 \\
FRB180814.J1554+74 & 3.0 & 9.0 & 0.77 & 3.04 & * & 1.00 \\
FRB180817.J1533+42 & 1.75 & -12.0 & 0.00 & 0.00 & * & 1.00 \\
FRB180923 & 1.75 & -12.0 & 0.00 & 0.00 & * & 1.00 \\
FRB181016 & 2.75 & 8.0 & 0.53 & 1.96 & 0.14 & 1.00 \\
FRB181017 & 3.0 & 12.0 & 0.71 & 5.23 & 0.01 & 1.00 \\
FRB181228 & 3.0 & -11.0 & 1.30 & 3.56 & 0.16 & 1.00 \\
\enddata
\tablecomments{Column description: (1) Source name; (2) $\Gamma$ index of the power law which gave the maximum TS; (3) Time difference between the radio event time specified in the FRB catalog and the central time of the bin; (4) Flux upper limit (95\% confidence level) corresponding to the maximum TS found in photons per square centimeter  per second; (5) Maximum TS found for this source. (6) Gamma-ray to radio luminosity ratio (upper limit). Radio luminosity lower limit extracted from the FRB Catalog. (7) p-value for post trial correction taking into account all inspected sources, indices and time bins. (*) Radio flux in FRB catalog \citep{petroff16} appears as null. (**) Source fit did not converge due to low photon count.}
\end{deluxetable}

\clearpage
\end{landscape}

\newpage
\clearpage
\begin{landscape}

\begin{deluxetable}{cccccccc}
\tablecaption{Fanaroff-Riley type 0 radio galaxies:  summary of the parameters for the maximum TS.
\label{tab:FR0s}}
\tablecolumns{9}
\tablenum{2}
\tablewidth{0pt}
\tablehead{
\colhead{Source} &
\colhead{Gamma} & 
\colhead{R. A.} &
\colhead{Dec} &
\colhead{Flux ($10^{-9}$)} & 
\colhead{Test}& \colhead{L$_\gamma$/L$_{radio}$}&
\colhead{p-value} \\
\colhead{Name} &
\colhead{Index ($\Gamma$)}& 
\colhead{dist. ($\degree$)} & \colhead{dist. ($\degree$)} & 
\colhead{(ph cm$^{-2}$ s$^{-1}$)} & \colhead{Statistic}&
\colhead{(10$^{4}$)}&
\colhead{(post trial)}
}

\startdata
SDSS\_J010852.48-003919.4 & 3.0 & -0.2 & -0.2 & 12.29 & 8.33 & 2.40 & 1.00 \\
SDSS\_J011204.61-001442.4 & 3.0 & -0.5 & -0.1 & 5.40 & 1.64 & 0.64 & 1.00 \\
SDSS\_J011515.78+001248.4 & 3.0 & -0.5 & 0.5 & 14.41 & 9.25 & 0.72 & 1.00 \\
SDSS\_J015127.10-083019.3 & 2.5 & -0.5 & 0.1 & 10.01 & 13.69 & 0.92 & 1.00 \\
SDSS\_J020835.81-083754.8 & 1.75 & 0.1 & -0.1 & 0.13 & 1.09 & 0.10 & 1.00 \\
SDSS\_J075354.98+130916.5 & 2.5 & 0.4 & -0.5 & 9.06 & 8.47 & 3.91 & 1.00 \\
SDSS\_J080716.58+145703.3 & 1.75 & -0.1 & -0.3 & 0.39 & 3.95 & 0.28 & 1.00 \\
SDSS\_J083158.49+562052.3 & 1.75 & -0.4 & -0.2 & 0.70 & 4.03 & 1.58 & 1.00 \\
SDSS\_J083511.98+051829.2 & 3.0 & 0.5 & -0.1 & 8.55 & 2.78 & 1.83 & 1.00 \\
SDSS\_J084102.73+595610.5 & 1.75 & 0.4 & 0.4 & 1.05 & 4.64 & 2.51 & 1.00 \\
SDSS\_J084701.88+100106.6 & 2.25 & 0.2 & -0.3 & 3.20 & 3.75 & 0.63 & 1.00 \\
SDSS\_J090652.79+412429.7 & 3.0 & 0.5 & -0.1 & 6.77 & 2.92 & 0.28 & 1.00 \\
SDSS\_J090734.91+325722.9 & 2.25 & -0.5 & -0.5 & 3.96 & 7.52 & 0.39 & 1.00 \\
SDSS\_J090937.44+192808.2 & 3.0 & 0.1 & -0.1 & 16.40 & 10.82 & 0.53 & 1.00 \\
SDSS\_J091039.92+184147.6 & 3.0 & -0.4 & -0.5 & 20.25 & 15.70 & 0.86 & 1.00 \\
SDSS\_J091601.78+173523.3 & 3.0 & -0.5 & 0.4 & 18.62 & 14.55 & 1.62 & 1.00 \\
SDSS\_J091754.25+133145.5 & 2.0 & 0.1 & -0.5 & 0.18 & 0.09 & 0.07 & 1.00 \\
\textbf{ SDSS\_J093003.56+341325.3 } & \textbf{ 3.0 } & \textbf{ 0.5 } & \textbf{ 0.5 } & \textbf{ 18.40 } & \textbf{ 18.60 } & \textbf{ 1.19 } & \textbf{ 0.72 }\\
SDSS\_J093346.08+100909.0 & 2.25 & -0.3 & 0.5 & 5.04 & 8.25 & 0.45 & 1.00 \\
SDSS\_J093938.62+385358.6 & 2.75 & -0.4 & 0.5 & 7.18 & 5.07 & 2.97 & 1.00 \\
SDSS\_J094319.15+361452.1 & 3.0 & -0.4 & -0.5 & 8.34 & 4.29 & 0.24 & 1.00 \\
SDSS\_J100549.83+003800.0 & 1.75 & 0.5 & -0.1 & 0.61 & 1.14 & 0.55 & 1.00 \\
SDSS\_J101329.65+075415.6 & 1.75 & 0.5 & -0.5 & 0.35 & 1.75 & 0.93 & 1.00 \\
SDSS\_J101806.67+000559.7 & 2.25 & -0.5 & -0.1 & 3.82 & 4.04 & 1.22 & 1.00 \\
SDSS\_J102403.28+420629.8 & 2.0 & -0.4 & -0.1 & 0.68 & 1.98 & 1.00 & 1.00 \\
SDSS\_J102511.50+171519.9 & 3.0 & 0.0 & -0.5 & 13.29 & 8.95 & 2.85 & 1.00 \\
SDSS\_J102544.22+102230.4 & 3.0 & 0.4 & 0.5 & 16.27 & 11.94 & 0.46 & 1.00 \\
\textbf{ SDSS\_J103719.33+433515.3 } & \textbf{ 3.0 } & \textbf{ -0.4 } & \textbf{ 0.5 } & \textbf{ 18.82 } & \textbf{ 26.61 } & \textbf{ 0.32 } & \textbf{ 0.02 }\\
SDSS\_J103952.47+205049.3 & 3.0 & -0.1 & -0.5 & 13.19 & 9.11 & 4.08 & 1.00 \\
SDSS\_J104028.37+091057.1 & 3.0 & 0.1 & 0.5 & 17.95 & 14.44 & 0.56 & 1.00 \\
SDSS\_J104403.68+435412.0 & 3.0 & 0.5 & -0.1 & 7.43 & 4.42 & 0.50 & 1.00 \\
\textbf{ SDSS\_J104811.90+045954.8 } & \textbf{ 3.0 } & \textbf{ -0.5 } & \textbf{ 0.5 } & \textbf{ 23.84 } & \textbf{ 21.20 } & \textbf{ 1.04 } & \textbf{ 0.28 }\\
SDSS\_J104852.92+480314.8 & 2.5 & -0.5 & -0.1 & 2.54 & 2.39 & 0.42 & 1.00 \\
SDSS\_J105731.16+405646.1 & 2.0 & 0.1 & 0.3 & 1.37 & 5.22 & 0.27 & 1.00 \\
SDSS\_J111113.18+284147.0 & 2.0 & -0.4 & 0.5 & 3.03 & 11.41 & 0.65 & 1.00 \\
SDSS\_J111622.70+291508.2 & 2.25 & -0.4 & 0.2 & 2.77 & 4.23 & 0.18 & 1.00 \\
SDSS\_J111700.10+323550.9 & 2.75 & 0.5 & -0.5 & 10.16 & 10.17 & 1.47 & 1.00 \\
SDSS\_J112029.23+040742.1 & 1.75 & -0.1 & 0.3 & 1.08 & 5.42 & 3.01 & 1.00 \\
SDSS\_J112039.95+504938.2 & 3.0 & 0.5 & -0.5 & 3.04 & 0.82 & 0.28 & 1.00 \\
SDSS\_J112256.47+340641.3 & 2.5 & -0.3 & 0.5 & 1.10 & 0.34 & 0.22 & 1.00 \\
SDSS\_J112625.19+520503.5 & 1.75 & -0.2 & -0.1 & 0.31 & 3.01 & 0.70 & 1.00 \\
SDSS\_J112727.52+400409.4 & 1.75 & 0.1 & 0.5 & 0.68 & 6.45 & 1.04 & 1.00 \\
SDSS\_J113446.55+485721.9 & 1.75 & 0.4 & -0.1 & 0.55 & 7.07 & 0.91 & 1.00 \\
SDSS\_J113449.29+490439.4 & 1.75 & 0.2 & -0.3 & 0.65 & 8.52 & 0.41 & 1.00 \\
SDSS\_J113637.14+510008.5 & 2.75 & -0.4 & -0.5 & 7.19 & 7.85 & 1.97 & 1.00 \\
SDSS\_J114230.94-021505.3 & 1.75 & -0.5 & 0.3 & 0.43 & 2.12 & 0.99 & 1.00 \\
SDSS\_J114232.84+262919.9 & 2.75 & 0.5 & 0.5 & 6.06 & 3.23 & 0.36 & 1.00 \\
SDSS\_J114804.60+372638.0 & 1.75 & 0.4 & -0.5 & 0.60 & 6.13 & 0.43 & 1.00 \\
SDSS\_J115531.39+545200.4 & 2.0 & -0.5 & 0.4 & 0.89 & 1.60 & 0.25 & 1.00 \\
SDSS\_J120551.46+203119.0 & 2.0 & -0.4 & -0.2 & 0.79 & 1.15 & 0.08 & 1.00 \\
SDSS\_J120607.81+400902.6 & 2.25 & 0.4 & 0.1 & 1.76 & 2.86 & 0.86 & 1.00 \\
SDSS\_J121329.27+504429.4 & 1.75 & -0.5 & -0.3 & 0.51 & 3.48 & 0.11 & 1.00 \\
SDSS\_J121951.65+282521.3 & 2.0 & 0.5 & -0.2 & 8.08 & 14.54 & 8.46 & 1.00 \\
SDSS\_J122421.31+600641.2 & 2.75 & 0.5 & 0.4 & 8.15 & 9.22 & 3.38 & 1.00 \\
SDSS\_J123011.85+470022.7 & 1.75 & 0.2 & -0.1 & 0.36 & 3.19 & 0.08 & 1.00 \\
SDSS\_J124318.73+033300.6 & 2.75 & 0.4 & 0.5 & 4.84 & 1.78 & 0.19 & 1.00 \\
SDSS\_J124633.75+115347.8 & 3.0 & -0.5 & -0.1 & 7.19 & 3.13 & 0.25 & 1.00 \\
SDSS\_J125027.42+001345.6 & 2.0 & -0.3 & 0.5 & 1.40 & 3.62 & 0.22 & 1.00 \\
SDSS\_J125409.12-011527.1 & 2.5 & -0.5 & 0.5 & 4.48 & 2.73 & 1.81 & 1.00 \\
SDSS\_J130404.99+075428.4 & 2.25 & 0.3 & -0.3 & 2.04 & 2.59 & 0.90 & 1.00 \\
SDSS\_J130837.91+434415.1 & 1.75 & 0.0 & -0.5 & 1.52 & 5.66 & 0.55 & 1.00 \\
SDSS\_J133042.51+323249.0 & 1.75 & -0.3 & 0.5 & 0.34 & 1.98 & 0.40 & 1.00 \\
SDSS\_J133455.94+134431.7 & 2.0 & -0.2 & 0.3 & 1.50 & 4.45 & 0.34 & 1.00 \\
SDSS\_J133621.18+031951.0 & 1.75 & -0.1 & 0.5 & 0.21 & 2.15 & 0.15 & 1.00 \\
SDSS\_J133737.49+155820.0 & 2.25 & -0.5 & -0.4 & 3.09 & 4.69 & 0.55 & 1.00 \\
SDSS\_J134159.72+294653.5 & 1.75 & -0.4 & 0.1 & 0.21 & 0.95 & 0.41 & 1.00 \\
SDSS\_J135036.01+334217.3 & 1.75 & -0.4 & 0.1 & 0.20 & 2.35 & 0.04 & 1.00 \\
\textbf{ SDSS\_J135226.71+140528.5 } & \textbf{ 2.0 } & \textbf{ 0.3 } & \textbf{ 0.5 } & \textbf{ 3.74 } & \textbf{ 19.01 } & \textbf{ 1.30 } & \textbf{ 0.64 }\\
SDSS\_J140528.32+304602.0 & 1.75 & -0.4 & 0.5 & 0.82 & 5.65 & 2.39 & 1.00 \\
SDSS\_J141451.35+030751.2 & 2.5 & -0.2 & -0.5 & 5.34 & 4.96 & 0.64 & 1.00 \\
SDSS\_J141517.98-022641.0 & 3.0 & -0.5 & 0.5 & 9.72 & 4.40 & 1.09 & 1.00 \\
SDSS\_J142724.23+372817.0 & 2.0 & -0.5 & -0.2 & 0.71 & 1.40 & 0.29 & 1.00 \\
SDSS\_J143156.59+164615.4 & 2.5 & -0.5 & -0.1 & 3.51 & 2.07 & 1.23 & 1.00 \\
SDSS\_J143312.96+525747.3 & 3.0 & -0.5 & 0.5 & 10.64 & 9.78 & 1.43 & 1.00 \\
SDSS\_J143424.79+024756.2 & 1.75 & 0.3 & -0.1 & 1.17 & 8.37 & 3.46 & 1.00 \\
SDSS\_J143620.38+051951.5 & 1.75 & 0.3 & 0.5 & 0.79 & 7.56 & 0.89 & 1.00 \\
SDSS\_J144745.52+132032.2 & 1.75 & 0.3 & -0.3 & 0.19 & 0.63 & 0.58 & 1.00 \\
SDSS\_J145216.49+121711.5 & 3.0 & 0.5 & -0.5 & 6.84 & 1.99 & 1.83 & 1.00 \\
SDSS\_J145243.25+165413.4 & 2.25 & 0.4 & -0.5 & 6.07 & 10.10 & 1.62 & 1.00 \\
SDSS\_J145616.20+203120.6 & 3.0 & -0.5 & -0.4 & 11.14 & 7.85 & 0.95 & 1.00 \\
SDSS\_J150152.30+174228.2 & 2.5 & -0.2 & 0.5 & 5.40 & 4.39 & 0.91 & 1.00 \\
SDSS\_J150425.68+074929.7 & 3.0 & 0.3 & 0.5 & 20.01 & 12.63 & 5.45 & 1.00 \\
\textbf{ SDSS\_J150601.89+084723.2 } & \textbf{ 3.0 } & \textbf{ -0.5 } & \textbf{ 0.5 } & \textbf{ 31.81 } & \textbf{ 25.24 } & \textbf{ 8.18 } & \textbf{ 0.04 }\\
\textbf{ SDSS\_J150636.57+092618.3 } & \textbf{ 3.0 } & \textbf{ -0.5 } & \textbf{ 0.5 } & \textbf{ 55.95 } & \textbf{ 60.47 } & \textbf{ 4.44 } & \textbf{5.8 $\times$ 10$^{\textbf{-10}}$}\\
SDSS\_J150808.25+265457.6 & 2.25 & 0.2 & 0.3 & 4.15 & 9.56 & 0.99 & 1.00 \\
SDSS\_J152010.94+254319.3 & 2.75 & 0.5 & -0.1 & 6.56 & 4.56 & 0.93 & 1.00 \\
SDSS\_J152151.85+074231.7 & 2.0 & 0.1 & -0.5 & 1.43 & 3.30 & 1.05 & 1.00 \\
SDSS\_J153016.15+270551.0 & 3.0 & 0.5 & -0.5 & 11.99 & 9.59 & 2.00 & 1.00 \\
SDSS\_J154147.28+453321.7 & 2.25 & -0.5 & -0.5 & 1.54 & 2.49 & 0.82 & 1.00 \\
SDSS\_J154426.93+470024.2 & 1.75 & -0.2 & -0.1 & 0.34 & 3.50 & 0.41 & 1.00 \\
SDSS\_J154451.23+433050.6 & 1.75 & -0.5 & 0.5 & 0.43 & 2.52 & 0.80 & 1.00 \\
\textbf{ SDSS\_J155951.61+255626.3 } & \textbf{ 3.0 } & \textbf{ -0.2 } & \textbf{ -0.5 } & \textbf{ 19.37 } & \textbf{ 24.00 } & \textbf{ 0.29 } & \textbf{ 0.07 }\\
SDSS\_J155953.99+444232.4 & 1.75 & -0.4 & 0.1 & 0.42 & 6.03 & 0.15 & 1.00 \\
SDSS\_J160426.51+174431.1 & 1.75 & 0.5 & 0.5 & 0.47 & 3.40 & 0.10 & 1.00 \\
SDSS\_J160523.84+143851.6 & 3.0 & -0.2 & -0.5 & 14.92 & 8.79 & 3.75 & 1.00 \\
SDSS\_J160616.02+181459.8 & 1.75 & 0.2 & 0.2 & 0.63 & 4.01 & 0.03 & 1.00 \\
SDSS\_J160641.83+084436.8 & 1.75 & -0.3 & -0.3 & 0.64 & 6.48 & 1.41 & 1.00 \\
SDSS\_J161238.84+293836.9 & 3.0 & -0.2 & -0.4 & 8.73 & 6.03 & 0.69 & 1.00 \\
SDSS\_J161256.85+095201.5 & 2.0 & -0.1 & -0.5 & 2.62 & 8.47 & 1.06 & 1.00 \\
SDSS\_J162146.06+254914.4 & 1.75 & -0.1 & 0.0 & 1.03 & 7.95 & 2.30 & 1.00 \\
SDSS\_J162549.96+402919.4 & 2.75 & -0.2 & 0.2 & 4.49 & 2.82 & 0.12 & 1.00 \\
SDSS\_J162846.13+252940.9 & 2.25 & 0.1 & 0.2 & 3.00 & 4.90 & 0.55 & 1.00 \\
SDSS\_J162944.98+404841.6 & 3.0 & 0.5 & -0.5 & 7.80 & 3.50 & 2.15 & 1.00 \\
SDSS\_J164925.86+360321.3 & 3.0 & -0.3 & -0.2 & 6.90 & 3.41 & 1.30 & 1.00 \\
SDSS\_J165830.05+252324.9 & 2.0 & 0.4 & 0.5 & 2.58 & 7.84 & 1.76 & 1.00 \\
SDSS\_J170358.49+241039.5 & 3.0 & -0.5 & -0.4 & 10.37 & 6.00 & 0.70 & 1.00 \\
SDSS\_J171522.97+572440.2 & 2.25 & -0.3 & -0.3 & 2.44 & 4.32 & 0.20 & 1.00 \\
SDSS\_J172215.41+304239.8 & 1.75 & -0.5 & 0.1 & 0.20 & 0.65 & 0.50 & 1.00 \\
\enddata

\tablecomments{Column description: (1) Source name; (2) (3) Distance added to the right ascension (declination) component of the source's position used in the fitting model; (4) $\Gamma$ index of the power law which gave the maximum TS; (5) Flux upper limit (95\% confidence level) corresponding to the maximum TS found in photons per square centimeter per second; (6) Maximum TS found for this source. (7) Gamma-ray to radio luminosity ratio. Radio luminosity and z extracted from the FR0 Catalog. (8) p-value for post trial correction taking into account all inspected sources, indices and spacial bins.}
\end{deluxetable}

\clearpage
\end{landscape}

\acknowledgments

S.B. acknowledges support from the Peruvian National Council for Science, Technology and Technological Innovation scholarship under grant CONCYTEC-FONDECyT 233-2015-2. J.B. acknowledges funding by the Direcci\'on de Gesti\'on de la Investigaci\'on at PUCP under grant No. DGI-2017-3-0019.  The authors would like to thank A. Gago and J. Jones, for useful suggestions and reading of the manuscript. We are also thankful to the reviewer for the comments and suggestions that helped us to improve this work. 


\end{document}